\documentstyle[aps,prb,multicol,epsfig,amsfonts]{revtex}

\begin{document}
\draft
\title{ Superconductivity in compressed iron: Role of spin 
fluctuations
}
\author{I.I. Mazin, D.A. Papaconstantopoulos, and M.J. Mehl}
\address{Center for Computational Materials Science,\\
Naval Research Laboratory, Washington, DC 20375-5000, USA}
\maketitle

\begin{abstract}
The recent discovery of superconductivity in hexagonal iron under
pressure poses a question about whether it is of conventional
(phonon) or unconventional (magnetic?) origin. We present 
first-principles calculations of the
electron-phonon coupling in iron at $P\agt 15$ GPa, and
argue that a conventional mechanism can explain the
appearance of  superconductivity, but not its rapid
disappearance at $P\agt 30$ GPa. We suggest that 
spin fluctuations, ferro- and/or antiferromagnetic,
play a crucial role in superconductivity in this case.
\end{abstract}

\begin{multicols}{2}
The recent report of  superconductivity in Fe under pressure \cite{fe} was 
among
 the latest news in the chain of discoveries of interesting cases of
superconductivity in conventional materials: MgB$_2$\cite{mgb2}, MgCNi$_3$\cite%
{mgcni3}, ZrZn$_2$\cite{zrzn2}, and others. At first glance, there is
nothing unexpected in the fact that iron, a transition metal, becomes
superconducting in a nonmagnetic phase. However, a closer look at this case
uncovers a number of strange facts which do not square with this simplistic
scenario. First, in the phase diagram (Fig.1) according to Ref. \onlinecite{fe}
the superconductivity appears at the same pressure ($P\approx 15$ GPa),
 at which the
bcc-hcp phase transition takes place. In other words, the bcc phase is not
superconducting (and is magnetic), while the hcp phase is already
 superconducting
at the lowest pressure at which it is stable. What is surprising,
however, is that the critical temperature is zero at the phase transition
pressure, and then grows gradually with pressure. If the superconductivity
were due to phonons, its appearence exactly at the crystallographic
phase transition would be just a coincidence, since there is no
reason for superconductivity to be suppressed near such a (first order)
phase transition. Second, the superconductivity disappears
very rapidly with increasing
pressure. Essentially, superconductivity occurs only in a narrow range of
compression, 132  Bohr$^3$$<V<145$ Bohr$^3$, where $V$ is the unit cell
volume (with two atoms per cell).
As we show below, neither electronic (e.g., density
of states) characteristics
 nor phonon frequencies change with pressure rapidly enough to reduce
the critical temperature from nearly 2 K to zero at such a small relative
compression.

On the other hand, there are cases in nature when superconductivity exists
only in a narrow pressure range, as in the recently discovered
ferromagnetic superconductor UGe$_{2}.$\cite{uge2} The accepted
explanation 
for such  behavior in this case is that superconductivity there
is believed to
be induced by spin fluctuations, which, first, stiffen rapidly away from the
quantum critical point (that is, the pressure that corresponds to the Curie
temperature $T_{C}=0),$ and, second, grow in magnitude close to this point.
Thus, the critical temperature goes to zero at the critical point, then
rises away from it, but disappears again when the energy of the
spin fluctuations becomes too high.  \cite{fay,kir}.

Finally, it is not even clear whether hcp iron is indeed nonmagnetic
over its whole range of stability. Density functional calculations using
the Generalized Gradient Approximation (GGA), which is usually very reliable
for 3d metals, clearly
show an antiferromagnetic ground state\cite{ron}, or a noncollinear magnetic
state\cite{somali}, but not a paramagnetic one. Furthermore, the calculated
elastic properties in magnetic hcp iron are in good agreement with  experiment,
while those for a nonmagnetic iron seriously disagree with experiment.%
\cite{ron} Admittedly, there is one experiment that is hard to explain if
hcp iron is magnetic even locally, namely  M\"{o}ssbauer spectroscopy
under pressure\cite{moss}. Given this contradiction, we take the point of
view that the question of magnetism in hcp iron is still open, and even
if the hcp iron is in fact non-magnetic, it should be very close to a magnetic 
instability.

In 1979, well before superconductivity was discovered in hcp iron,
Wohlfarth pointed out that 
spin-fluctuations may play a role.  \cite{wolf}
He noticed that at the lowest pressures at which the hcp iron is stable,
it must be very close to an antiferromagnetic instability, which should lead
to pair-breaking spin fluctuations. He also pointed out that these fluctuations
should disappear at higher pressure, as the density of states becomes smaller,
and thus at some pressure superconductivity is to be expected. Of course,
what is missing from this scenario is the fact that superconductivity
disappears again rather soon when pressure is increased.

In order to elucidate the potential for conventional, electron-phonon
superconductivity in iron we performed first-principles 
Linear Augmented Plane Wave (LAPW) calculations in 
GGA of the
following quantities: (a) the electron-ion matrix element, 
$W,$ (b) the density of
states at the Fermi level, $N,$ and (c) the Stoner factor, $I.$ 
Superconductivity in transition metals has been subject to numerous studies
in the 70's and 80's (see, e.g., Ref.\onlinecite{rainer} for review), and is
quantitatively understood. The McMillan formula, 
\begin{equation}
T_{c}=\frac{ \omega _{\log }}{1.2}\exp [\frac{-1.02(1+\lambda )}{%
\lambda -\mu ^{\ast }(1+0.54\lambda )}],  \label{mm1}
\end{equation}%
describes $T_{c}$ reasonably well. The parameters of the formula have
the following meanings: $\omega _{\log }$ is the logarithmically averaged
phonon frequency, $\mu ^{\ast }$ is the Coulomb pseudopotential, which for
transition metals is usually close to 0.1,
and $\lambda $ is the electron-phonon
coupling constant, $\lambda =\eta M^{-1}\left\langle \omega
^{-2}\right\rangle ,$ where $\eta =NW$ is the so-called Hopfield parameter, $%
M$ is the ion mass, and $\left\langle \omega ^{-2}\right\rangle $ is the
average inverse square of the phonon frequency. Note that the electron-phonon
coupling is responsible for both pairing interaction and mass
renormalization. The constant $\lambda $ measures both effects and appears
three times in the McMillan formula: once as a measure of the pairing
strength, and twice (in $1+\lambda $ and in $1+0.54\lambda )$ because of
mass renormalization.

In at least one transition metal, Pd, the electron-phonon coupling is
definitely strong enough to render the material superconducting but, 
because it is near a
magnetic instability, superconductivity is suppressed.
 In other words, charge fluctuations
(phonons) coexist in Pd with spin-fluctuations (paramagnons), which have
a pair-breaking effect for s-wave symmetry, but contribute to the mass
renormalization like phonons. These effects can be approximately accounted
for by modifying  the McMillan formula as follows:%
\begin{equation}
T_{c}=\frac{ \omega _{\log }}{1.2}\exp \{\frac{-1.02(1+\lambda
_{ph}+\lambda _{sf})}{\lambda _{ph}-\lambda _{sf}-\mu ^{\ast
}[1+0.54(\lambda _{ph}+\lambda _{sf})]}\},  \label{mm2}
\end{equation}%
where the subscripts stand for ''phonons'' and ''spin fluctuations'', so
that $\lambda _{sf}$ is an electron-paramagnon coupling constant.

Since the Migdal theorem does not hold for paramagnons, it is hardly
possible to derive
 a formula for $\lambda _{sf}$ and compute it from
first principles. Within the framework of the Stoner model, and near the
magnetic instability, one can estimate coupling with ferromagnetic
spin fluctuations as\cite{fay}%
\begin{equation}
\lambda _{sf}\propto\left\langle \frac{1}{1-NI}\right\rangle
\approx \frac{\alpha 
}{1-NI}, \label{ston}
\end{equation}
where $I$ is the Stoner factor and $\alpha $ characterizes the band
structure and the $q-$dependence of the bare magnetic susceptibility\cite%
{note}. The most accurate way to estimate $I$ is by performing fixed spin
moment calculations for small moments and taking the second derivative
of the total energy with respect to the fixed moment, $I=N^{-1}-2d^{2}E/dm^{2}.$
With this method we calculated $I=$0.075 Ry/atom, using
 fixed-spin-moment LAPW total
energies at $V=146.53$ Bohr$^3$ . Note that $I$ is practically pressure-independent.

For {\it antiferromagnetic} spin fluctuations Eq.\ref{ston} is not applicable,
but instead one can write, as a  first approximation, \[
\lambda _{sf}\propto\left\langle \frac{1}{1-\chi I}\right\rangle
\approx \frac{\alpha
}{1-\chi I}, 
\]%
where $\chi$ is the noninteracting spin susceptibility at the wave vector
corresponding to the antiferromagnetic ordering. The main effect of pressure
on the electronic structure of  hcp iron is an overall scaling of the 
bandwidth. Thus, one can assume that $\chi\approx \beta N$,
where $\beta$ is a pressure independent constant.  

All quantities entering Eq. \ref{mm2}, except $\alpha$ and $\beta ,$ were
calculated from first principles, using the approximations outlined above
with the phonon frequencies, calculated elsewhere
\cite{phonons-t}, which agree well with experiment\cite{phonons}.
It is worth noting that because of all these approximations we do not
expect to obtain quantitatively the value for $T_{c},$ but we do expect to
describe correctly the trends associated with the pressure\cite{soft}.

In the calculations of the electron-phonon Hopfield parameter $\eta$,
we used the "rigid-muffin-tin" theory of Gaspari and Gyorffy.\cite{GG}
According to this theory, which is known to work quite well for
transition metals (see, e.g. Ref.\onlinecite{papa}),
 $\eta=\sum_l{W^2_{l}N_lN_{l+1}/N}$, where 
$W_l$ are computed from particular
 integrals involving the radial wave functions,
as defined in Ref.\onlinecite{GG}, and $N_l$
is the partial density of states at the Fermi level with angular momentum 
$l$. We used self-consistent MT potentials with touching MT spheres,
calculated by
the LAPW method. It is well known\cite{papa}
 that for transition metals the main 
contribution to $\eta$ comes from $l=2$, thus $\eta\approx
W^2_{d}N_dN_f/N\approx W^2_{d}N_f$,
as the ratio $N_d/N$ is close to unity in transition metals. In
this connection, it is worth noting that
the volume dependence of $\eta$ (cf. Fig.\ref{eta}) is totally
different from that of the total density of states $N$; in fact
it has the  opposite variation.

The pressure-independent constants $\alpha$ and $\beta$ were
adjusted so as to have the magnetic instability ($\beta NI=1$)
near the hcp-bcc phase boundary, $V\approx 145$ Bohr$^3$, and to 
have the maximal critical temperature of the right order.
We used $\alpha=0.13 $ and $\beta=1.215$.
Qualitative dependence of $T_c$ on pressure is not sensitive
to the actual values of these constants.

\begin{figure}[tbp]
\centerline{\epsfig{file=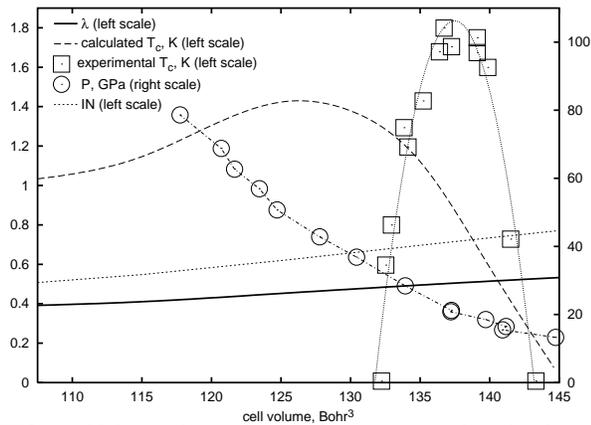,width=0.95\linewidth}}
\setlength{\columnwidth}{3.2in} \nopagebreak
\caption{Volume dependence of experimental and calculated superconducting 
properties of hcp iron. The electron-phonon coupling constant,
$\lambda$, and the critical temperature for the s-wave
superconductivity, $T_c$,
(according to Eq.\ref{mm2})
were computed as described in the text. The constant 
$\alpha$, controlling the strength of the pair-breaking effect
of spin fluctuations, was adjusted so as to have superconductivity
disappear at $V\approx 145$  Bohr$^3$, as in the experiment. The experimental
critical temperatures from Ref.
\protect\onlinecite{fe} are also plotted. For translating
the pressure scale into the volume scale we used the experimental equation
of state\cite{jep}, which is also shown in the Figure. We also
plotted the Stoner product, $IN$.}
\label{fig}
\end{figure}

\begin{figure}[tbp]
\centerline{\epsfig{file=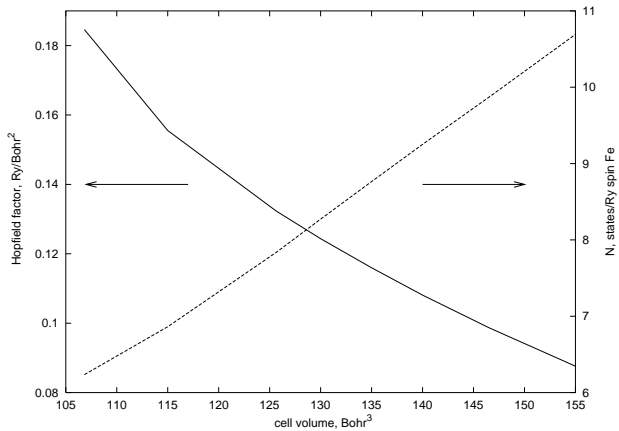,width=0.95\linewidth}}
\setlength{\columnwidth}{3.2in} \nopagebreak
\caption{Hopfield factor $\eta$ in Ry/Bohr$^2$ compared with
the density of states at the Fermi level, $N$.}
\label{eta}
\end{figure}

The results of our calculations are summarized in Table 1, 
and compared with  experiment on Fig.\ref{fig}. We observe
that the phase diagram can be described qualitatively within the framework
of the conventional scenario (pairing phonons plus pair-breaking spin
fluctuations), in the sense that the critical temperature rises with pressure
near
 the bcc-fcc transition pressure, reaches a maximum, and
then decays. The underlying physics is as follows: spin fluctuations,
which get stronger with lowering pressure, suppress superconductivity,
and at very high pressures the lattice stiffens and $T_c$
goes down. However, the range of pressures at which superconductivity
exists is much larger in the calculations. This is not accidental; the only
mechanism that prevents superconductivity at higher pressure is stiffening
of the lattice, since the electronic part, $\eta$, actually {\it grows}
with pressure (Fig.\ref{eta}).
 Indeed,  hcp Fe does stiffen rapidly under pressure.
However, it is by far not strong enough an effect to destabilize superconductivity at $%
P>35$ GPa (corresponding to the atomic volume of 132  Bohr$^3$, which is
only a 4\%
compression from the maximum $T_{c}$ volume). In other words, elastic
properties do not change fast enough with pressure to explain such an
extraordinarily narrow range of superconductivity.

\begin{figure}[tbp]
\centerline{\epsfig{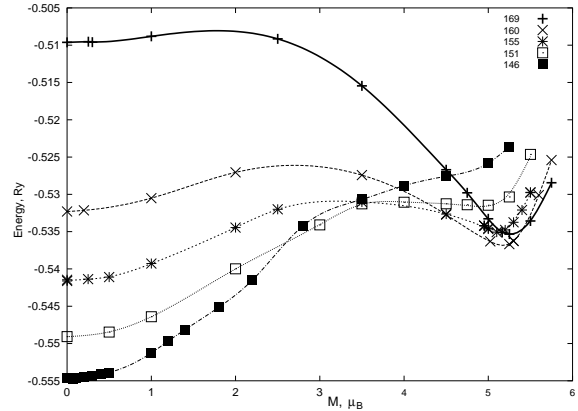}}
\setlength{\columnwidth}{3.2in} \nopagebreak
\caption{LAPW fixed spin moment calculations for ferromagnetic hcp
iron. Magnetic moments and energies are given on a per cell (two
atoms) basis. The curves correspond,
from up down, to the volumes per atom: 169.97  Bohr$^3$,
160.44  Bohr$^3$, 155.81  Bohr$^3$, 151.27  Bohr$^3$, and 146.83  Bohr$^3$.
}
\label{fixed}
\end{figure}

On the other hand, {\it magnetic} properties, at least as calculated within
density functional theory, {\it do change} in the relevant range.
Indeed, at $V=180$  Bohr$^3$, both antiferromagnetic arrangements considered in
Ref.\onlinecite{ron}
are stable and produce the same magnetic moment, on the order of  2 $%
\mu _{B} $ per Fe atom\cite{errata}.
 At $V=145$  Bohr$^3$, the largest volume at which
hcp iron exists, both
arrangements are still stable, and the corresponding magnetic moments are
similar, although they reduce to values of
$\approx 0.5$ $\mu _{B}.$ One can interpret this
as local magnetism on Fe with nearest-neighbor antiferromagnetic exchange ($%
J).$ The exchange energy for the magnetic structure afmI (we use the
notation of Ref.\onlinecite{ron}) is zero, for afmII
it is -2$J$ per atom, and for a ferromagnetic state it is $+6J$ per atom.
Apparently, the latter energy is sufficiently large to render the
ferromagnetic state unstable. In fact, and this has not been noticed before,
while the ferromagnetic state is energetically unfavorable compared
with the nonmagnetic state
for all volumes up to $V=157$  Bohr$^3$, there is a metastable ferromagnetic state
(metamagnetism) with $M\approx 2.6$ $\mu_B$/atom, for all $V>145$  Bohr$^3$ 
(Fig.\ref{fixed}) On the other hand, while antiferromagnetic states,
stabilized by the exchange energy, are lower in energy than the nonmagnetic one,
the equilibrium magnetization is rather small, on the order of 1 $\mu_B$. 
We find, in agreement with
Ref.\onlinecite{ron},
that the antiferromagnetism disappears only at $V\approx 120$  Bohr$^3$. 

While the picture derived from these calculations is at odds with that deduced
from the M\"ossbauer experiment, in the sense that the calculated tendency to 
magnetism appears to be too strong, the main conclusion can be expressed
qualitatively as follows:
the spectrum of the spin fluctuations changes with
pressure rapidly and drastically near $P\approx 20$ GPa, and probably in a very
nontrivial way, reflecting two different types of magnetic 
excitations: antiferromagnetic, and also ferromagnetic (the latter, possibly,
 of highly
 nonlinear metamagnetic character). 
 This strongly suggests
an exclusive role played by spin fluctuations in iron
 superconductivity around $%
P\approx 20$ GPa. Whether the superconductivity is s-wave, induced by the
electron-phonon interaction, with spin fluctuations being ordinary pair
breakers, or it is d-wave or p-wave, with
spin fluctuations being the pairing agents, cannot be firmly established
now, and should be
clarified by further experimental studies. However, the latter scenario
seems to be
not only more exciting, but also more likely, if we believe that, according to
the M\"ossbauer results,
there are no localized magnetic moments in the whole superconductivity
range, because this means that the spin fluctuations become weaker away from
the bcc-hcp transition, and their pair-breaking effect can only diminish with
pressure. On the other hand, if superconductivity is d- or p-wave, and the
spin fluctuations are pairing, the phase diagram looks consistent with the
standard prediction for this scenario\cite{fay}.
A key experiment in this situation would be
measuring  the effect of nonmagnetic impurities on
superconductivity.

This work was supported by  the Office of Naval Research. The authors
are grateful to D.J. Singh, A. F. Goncharov and R.E. Cohen for useful
discussions.


\begin{table}
\caption{Calculated parameters at three representative volumes.}
\begin{tabular}{llll}
$V,$  Bohr$^3$ & 106.83 &   125.63   & 146.53 \\ 
$P$, GPa$^{(a)}$ & 79   & 48   & 12 \\ 
$\eta ,$ Ry/Bohr$^{2}$ & 0.185   & 0.132   & 0.099 \\ 
$1/\sqrt{\left\langle \omega ^{-2}\right\rangle }^{{}},$ cm$^{-1 (b)}$
& 332 &   260 &   206 \\ 
$\omega _{\log },$ cm$^{-1 (b)}$ & 343   & 260   & 200 \\ 
$N,$ st./Ry atom spin & 6.7   & 8.3 &   10.4 \\ 
$1/(1-IN)$ & 2.0 &   2.7  & 4.6 \\ 
$T_{c},$ K$,$ Eq.\ref{mm1} & 2.5 & 4.1 & 6.0   \\ 
$T_{c},$ K$,$ Eq.\ref{mm2} ($\alpha =0.013, \beta=1.215)$  & 1.0 &1.4  & 0.0 \\ 
\end{tabular}
$^{(a)}$Ref.\onlinecite{jep}\\
 $^{(b)}$These values are calculated from the data in
Ref.\onlinecite{phonons-t}, assuming that the
phonon density of states is uniformly distributed
between the longitudinal and the transverse peaks. In fact, any reasonable
assumption about the shape of the phonon DOS leads to very similar results.
\end{table}
\end{multicols}

\begin{references}
\bibitem{fe}K. Shimizu, T. Kimura, S. Furomoto, K. Takeda, K. Kontani, Y. Onuki,
and K. Amaya, Nature (London), {\bf  412}, 3169 (2001)


\bibitem{mgb2} J. Nagamatsu, N. Nakagawa, T. Muranaka, Y. Zenitani and J.
Akimitsu, Nature (London) {\bf 410}, 63 (2001).
\bibitem{mgcni3}
 T. He, Q. Huang, A.P. Ramirez, Y. Wang, K.A. Regan, N. Rogado, M.A. Hayward,
 M. K. Haas, J.S. Slusky, K. Inumaru, H.W. Zandbergen, N.P. Ong, and
R.J. Cava, cond-mat/0103296.

\bibitem{zrzn2}C. Peiderer, M. Uhlarz, S.M. Hayden, R. Vollmer,
H.v.Lohneysen, N.R. Bernhoeft and C.G. Lonzarich, Nature {\bf 412}, 58
(2001).

\bibitem{uge2}S. S. Saxena et al., Nature (London) 406, 587 (2000).
\bibitem{fay} D. Fay and J. Appel, Phys. Rev. {\bf B 22}, 3173 (1980).
\bibitem{kir} T.R. Kirckpatrick, D. Belitz, T. Vojta, and R. Narayanan,
cond-mat/0105627 (2001).
\bibitem{ron}G. Steinle-Neumann, L. Stixrude, and R.E. Cohen, 
Phys. Rev. {\bf B 60}, 791 (1999).
\bibitem{somali} R.E. Cohen, S. Gramsch, S. Mukherjee, G. Steinle-Neumann, L. Stixrude, cond-mat/0110025.

\bibitem{moss}R. D. Taylor, M. P. Pasternak, and R. Jeanloz, J. Appl.
 Phys. 69, 6126 (1991). 
\bibitem{wolf} E.P. Wohlfarth, Phys. Lett. {\bf 75A}, 141 (1979).
\bibitem{rainer}D. Rainer, {\it
Principles of ab initio calculations of superconducting transition temperatures}, 
in Progress in Low Temperature Physics, Vol. X, p. 373 (North Holland, 1986). 


\bibitem{note} The density functional calculations suggest that the spin
fluctuations should have substantial $q-$dependence with a maximum away from
the zone center. This follows from the fact\cite{ron}, that while
ferromagnetic ground state is unstable for the whole pressure range of
interest, various antiferromagnetic states are stable in the calculations.
\bibitem{phonons-t}D. Alfe, G.D. Price, and M.J. Gillan, 
Phys. Rev. {\bf B 64}, 045123 (2001); P. S\"oderlind, J.A. Moriarty, and 
J.M. Wills, Phys. Rev. {\bf B 51}, 14063 (1996).
\bibitem{phonons} H. Olijnyk, A.P.Jephcoat, and K. Refson, Europhys. Lett.
{\bf 53}, 504 (2001); 
\bibitem{soft} All published calculations and experiments
give roughly the same volume dependence for individual phonon
modes and for average phonon frequencies. Generally speaking, one may think about
 a phonon mode, which becomes soft at the bcc-hcp phase transition, and
is responsible for superconductivity. This could, in principle,
explain the narrow range of the existence of the superconducting phase.
However, the calculated dispersion curves (Ref. \onlinecite{phonons-t}) do
not show any unusual mode softening. We also explicitely checked
the most likely candidate for a soft mode, the in-plane optical
phonon, by performing LAPW frozen phonon calculations, 
and found that it is not particularly soft at the hcp-bcc
(273 cm$^{-1}$) and hardens with pressure with a normal Gr\"uneisen
parameter of 1 (at $V=130$  Bohr$^3$) to 1.7 (at $V=146$  Bohr$^3$).
\bibitem{GG}
 G.D. Gaspari and B.L. Gyorffy Phys. Rev. Lett. {\bf 28}, 801 (1972).
\bibitem{papa}D.A. Papaconstantopoulos, L.L. Boyer, B.M. Klein, A.R.
Williams, V.L. Moruzzi, J.F. Janak, 
Phys. Rev. {\bf B 15}, 4221 (1977).
\bibitem{errata} Note that in the Ref.\onlinecite{ron} the moments were
plotted incorrectly smaller by a factor of two\cite{somali}.
\bibitem{jep} A. Jephcoat, H. K. Mao, and P. Bell, J. Geophys. Res. {\bf 91},
4677 (1986); H. K. Mao, J. Xu, V. V. Struzhkin, J. Shu, R. J. Hemley, W. Sturhahn,
 M. Y. Hu, E. E. Alp, L. Vocadlo, D. Alfe, G. D. Price, M. J. Gillan, M.
Schwoerer-B\"ohning, D. H\"ausermann, P. Eng, G. Shen, H. Giefers, R. Lübbers, G. Wortmann,
Science, {\bf 292}, 914 (2001).
\end{references}
\end{document}